\documentclass[pdftex]{article}
\usepackage{spconf}
\usepackage{amsmath,amsthm,amsfonts,url,epsfig,placeins,xcolor,enumitem}
\usepackage[ruled,vlined,linesnumbered]{algorithm2e}
\usepackage{tikz}
\usepackage{caption}
\usepackage{subdepth}
\usepackage{bm}
\usepackage{accents}
\usepackage{wrapfig}
\usepackage{subfig}
\usepackage{hyperref}
\usetikzlibrary{calc,trees,positioning,arrows,chains,shapes.geometric,%
    decorations.pathreplacing,decorations.pathmorphing,shapes,%
    matrix,shapes.symbols}
\usepackage[capitalise]{cleveref}
\usepackage[export]{adjustbox}% http://ctan.org/pkg/adjustbox

\newcommand\copyrighttext{%
  \footnotesize \textcopyright Copyright 2021 IEEE. Published in ICASSP 2021 - 2021 IEEE International Conference on Acoustics, Speech and Signal Processing (ICASSP), scheduled for 6-11 June 2021 in Toronto, Ontario, Canada. Personal use of this material is permitted. However, permission to reprint/republish this material for advertising or promotional purposes or for creating new collective works for resale or redistribution to servers or lists, or to reuse any copyrighted component of this work in other works, must be obtained from the IEEE. Contact: Manager, Copyrights and Permissions / IEEE Service Center / 445 Hoes Lane / P.O. Box 1331 / Piscataway, NJ 08855-1331, USA. Telephone: + Intl. 908-562-3966.
}
\newcommand\Copyrightnotice{%
\begin{tikzpicture}[remember picture,overlay]
\node[anchor=south,yshift=10pt] at (current page.south) {\fbox{\parbox{\dimexpr\textwidth-\fboxsep-\fboxrule\relax}{\copyrighttext}}};
\end{tikzpicture}%
}
\SetAlgoCaptionSeparator{}

\graphicspath{ {Figures/} }

\newcommand{\C}{\mathbb{C}}

\newcommand{\R}{\mathbb{R}}

\newcommand{\allapps}{\iffalse}
\title{SiML: Sieved Maximum Likelihood for Array Signal Processing}
\name{Matthieu Simeoni $^{\ast}$, Paul Hurley $^\dagger$\thanks{Part of this work was carried out at the IBM Zurich Research Laboratory, where both authors previously worked. The work of Matthieu Simeoni was supported by the Swiss National
Science Foundation under Grant 200 021 181 978/1, “SESAM - Sensing and
Sampling: Theory and Algorithms”.}
}
\address{$^\ast$ \'Ecole Polytechnique F\'ed\'erale de Lausanne (EPFL), CH-1015 Lausanne, Switzerland\\
$^\dagger$ Western Sydney University (WSU), NSW-2150 Sydney, Australia}
\begin{document}
\ninept 

\maketitle
\Copyrightnotice

% 100-150 word abstract
\begin{abstract}
 \emph{Stochastic Maximum Likelihood (SML)} is a popular \emph{direction of arrival (DOA)} estimation technique in array signal processing. It is a parametric method that jointly estimates signal and instrument noise by maximum likelihood, achieving excellent statistical performance. Some drawbacks are the computational overhead as well as the limitation to a point-source data model with fewer sources than sensors.
In this work, we propose a \emph{Sieved Maximum Likelihood (SiML)} method. It uses a general functional data model, allowing an unrestricted number of arbitrarily-shaped sources to be recovered. To this end, we leverage functional analysis tools and express the data in terms of an infinite-dimensional sampling operator acting on a Gaussian random function. We show that SiML is computationally more efficient than traditional SML, resilient to noise, and results in much better accuracy than spectral-based methods. \end{abstract}

%, radio astronomy.}

\section{Introduction}
Array signal processing \cite{krim1996two,mailloux2005phased,johnson1992array} is primarily concerned with the sensing, processing and estimation of random wavefields (electromagnetic or mechanic). Techniques from array signal processing are used in myriad of applications, including for example  acoustics \cite{brandstein2013microphone,benesty2008microphone}, radio-interferometry \cite{thompson2008interferometry,simeoni2015towards}, radar and sonar systems \cite{mailloux2005phased,haykin1985array}, wireless networks \cite{godara1997application,paulraj1997space,hurley2016flexibeam},  and medical imagery \cite{zhi2000principles,rafaely2015fundamentals}. The sensing devices in all those applications consist of large networks of sensors, called \emph{sensor arrays} or \emph{phased-arrays}. 
%Such arrays come in various forms, ranging from cylindrical arrays of magnetic coils in functional MRI, to interferometers in radio-astronomy or microphone arrays in acoustics.    

A common task in array signal processing consists of estimating the intensity field of an emitting wavefield. The various algorithms available in the literature for this purpose divide into two categories~\cite{krim1996two}: \emph{spectral-based} and \emph{parametric} methods. 
Spectral-based
%, which encapsulate both steered-response power and subspace methods,
ones estimate the intensity field by ``steering" the array towards particular directions in space and evaluating the output power. The intensity field is thus recovered sequentially by scanning via \emph{beamforming} \cite{krim1996two,haykin1985array,hurley2016flexibeam} a grid covering the field of view. 
% so as to form a focused radiation pattern, able to isolate contributions from particular directions to the wavefield.
Famous beamformers include \emph{Bartlett}, also known as \emph{Matched Beamforming (MB)}, and \emph{Capon}, which is often called \emph{Minimum Variance Distortionless Response (MVDR)}~\cite{krim1996two}. These are extremely simple to implement, computationally attractive, and quite generic with little structural or distributional assumptions on the sensed wavefield. They are however limited in terms of accuracy, in particular for extreme acquisition conditions such as small sample size, low signal-to-noise ratio and/or spatial correlation in the wavefield \cite{krim1996two}. 

Parametric methods, on the other hand, attempt to overcome those limitations using a statistical model for the instrument noise and the unknown wavefield. Typically, the thermal noise at each sensor is modelled as an additive Gaussian white random process while the wavefield is assumed to be the incoherent sum of $Q$ point sources with amplitudes distributed according to a multivariate complex Gaussian distribution, where $Q$ is strictly smaller than the total number of sensors composing the array. In this context the problem of estimating the intensity field is generally referred to as a \emph{direction of arrival (DOA)} estimation problem \cite{krim1996two}. 
%, since from the structural assumptions on the intensity field only $Q$ directions in the field are nontrivial.
By thus specifying the data model, parametric methods can achieve much better recovery performance, both theoretically and in practice \cite{krim1996two}. \emph{Stochastic Maximum Likelihood (SML)} \cite{krim1996two,stoica1996maximum,stoica2001stochastic} is perhaps the most well-known parametric method. It uses explicit maximum likelihood expressions \cite{stoica1995concentrated} to estimate the various parameters involved in the traditional \emph{point-source model} \cite{krim1996two,stoica1996maximum}, namely the noise and sources power as well as the directions of arrival. Each parameter is estimated consistently and general theory of maximum likelihood  guarantees efficient estimation as the number of samples grows to infinity. 
These strong theoretical guarantees and excellent empirical performance come however at the cost of very intense computation~\cite{krim1996two}. Moreover, the data model assumptions restrict its application to point sources, of which there must be fewer than the number of sensors, preventing its use in many applications. 
This is particularly true for radio-astronomy \cite{thompson2008interferometry,simeoni2015towards}, where the number of sources is typically far larger than the number of antennas forming the array. Moreover, the increased resolution of modern radio-interferometers \cite{van2013lofar,dewdney2009square} permits celestial sources with spatial extent and complex shapes to be resolved, for which a point-source model is overly  simplistic. \\[1em]
The present work, SiML, takes a maximum likelihood approach based on more general \emph{functional} data model,  which, in particular, allows potentially correlated sources with spatial extent and arbitrary shapes to be recovered.
To this end, we leverage functional analysis tools \cite{vetterli2014foundations,ramsay2006functional} and formulate the data in terms of an infinite-dimensional \emph{sampling operator} \cite{vetterli2014foundations} acting on the wavefield's \emph{amplitude function}, modelled as a complex Gaussian random function \cite{lifshits2012lectures}. Based on this data model, we derive a joint maximum likelihood estimate for both the covariance kernel of this random function and the sensor noise power. 
%The estimate for the covariance kernel is non-parametric, described as a projection over a finite-dimensional space.
%As in non-parametric regression \cite{ramsay2006functional}, the covariance kernel is assumed to be representable  as a finite sum of functions.
As the optimisation problem elicits many solutions, we deploy the \emph{method of sieves} \cite{grenander1981abstract,geman1982nonparametric}  as a means of restricting the optimisation problem to a lower dimensional subspace.
%constrain the kernel search-space to a finite-dimensional subspace, spanned by some family of functions, sufficiently ``coherent" with the sensing device (in a sense that will be made precise).  
For identifiability, we show that this subspace must have a smaller dimension than the total number of sensors in the array.  
A suitable subspace dimension is obtained by trading off between likelihood improvement and model complexity through the \emph{Bayesian Information Criterion (BIC)} \cite{bhat2010derivation}. Simulations reveal that the subspace dimension acts as a regularisation parameter, increasing or decreasing with the SNR. 
%Using asymptotic arguments, a statistical $\chi^2$-test is derived in order to assess the optimal rank of the recovered covariance operator. 
For a known noise level, the resulting estimate of the covariance field is shown to be an unbiased, consistent and asymptotically efficient estimate of a particular oblique projection of the true covariance field.  The method is computationally far more efficient than traditional SML and resilient to noise. Finally, we demonstrate by simulation that  SiML obtains much better accuracy and contrast than spectral-based methods.

\section{A Functional Data Model}

To allow for the handling of very general sources, we introduce in this section a \emph{functional} data model. Leveraging tools from functional analysis \cite{vetterli2014foundations,ramsay2002applied}, the sensing device can be modelled as an infinite dimensional sampling operator acting on an unknown random \emph{amplitude function} (see \cite{lifshits2012lectures} for an introduction to random functions). We first investigate the \emph{population} version of the data model, where the covariance matrix of the instrument recordings is known,  before presenting its \emph{empirical} counterpart, where the covariance matrix is estimated from i.i.d. observations. More details on the modelling assumptions can be found in  \cite{krim1996two,stoica1996maximum,ottersten1998covariance,johnson1992array}.
\subsection{Population Version}
Consider an array of $L$ sensors with positions $\{\bm{p}_1,\ldots,\bm{p}_L\}\subset\R^3$.
%For simplicity, we first examine the case of a single physical source in the \emph{far-field} of the array, emitting a \emph{narrowband} waveform with frequency $f\in\R$ and \emph{deterministic} amplitude $s_q\in\C$. With these assumptions the signals $y_i$ received by the antennas only differ by a phase shift, leading to a planar wavefront of the form:
%$$ y_i(t)=s_q\;\mbox{exp}\left\{j2\pi f\left(t-\frac{\langle \bm{r}_q,\bm{p}_i\rangle}{c}\right)\right\}, \quad \forall t\in \R,$$
%where $\bm{r}_q\in\mathbb{S}^2$ denotes the direction of the emitting source and $c~\in~\R$ denotes the speed of light. 
Assuming the emitting sources are in the far-field \cite{johnson1992array} of the sensor array, they can be thought of as lying on the unit sphere $\mathbb{S}^2$. To allow for arbitrary numbers of complex sources, we consider a notional continuous \emph{source field} covering the entire sphere, with an associated \emph{amplitude function}, describing for each direction $\bm{r}\in\mathbb{S}^2$ the emission strength of the source field. In practice, source amplitudes fluctuate randomly \cite{krim1996two,stoica1996maximum}, and the amplitude function can  be modelled as a complex random function $\mathcal{S}=\{S(\bm{r}):\Omega\rightarrow \C,\bm{r}\in\mathbb{S}^2\}$, 
%\begin{align*}S:\begin{cases}\Omega\times \mathbb{S}^2&\rightarrow \C,\\
%(\omega,\bm{r})&\mapsto s_{\omega}(\bm{r}),
%\end{cases}
%\end{align*}
%$\omega$ an event of this probability space and $s_{\omega}:\mathbb{S}^2\rightarrow \C$ a \emph{sample function} \cite{lifshits2012lectures} of the random function $S$.
where $\Omega$ is some probability space.  More precisely, we assume $\mathcal{S}$ to be a \emph{Gaussian random function}~\cite{lifshits2012lectures}, i.e., that all its finite marginals have distribution: 
$$ \left[S(\bm{r}_1),\cdots,S(\bm{r}_n)\right]\stackrel{d}{\sim}\mathbb{C}\mathcal{N}_n(0,\mathcal{K}_{\mathcal{I}_n}),\,\forall \,\mathcal{I}_n\subset\mathbb{S}^2, \, \forall \,n\in\mathbb{N},$$
where $\mathbb{C}\mathcal{N}_n$ denotes the $n$-variate \emph{centrally symmetric, complex Gaussian distribution} \cite{goodman1963statistical,gallager2008principles}, $\mathcal{I}_n:=\{\bm{r}_1,\ldots,\bm{r}_n\}$ and $\mathcal{K}_{\mathcal{I}_n}\in\C^{n\times n}$ is some valid covariance matrix depending on the set ${\mathcal{I}_n}$.

%Assuming a far-field context, we model the complex amplitude distribution of the emitting wavefield as a random \emph{aperture function} $S:\Omega\times \mathbb{S}^2\rightarrow \C$ defined over the sphere, where $\Omega$ denotes some probability space. This formalism is particularly convenient as it is agnostic on both the shape and number of physical  sources in the emitting wavefield, considering it as a continuous entity with a potential source in every direction. As discussed in \cite{krim1996two}, it is reasonable to assume a centred, centrally symmetric, complex Gaussian distribution for the sources' amplitude 
From the Huygens-Fresnel principle \cite{mast2007fresnel}, exciting the source field with a narrowband waveform of wavelength $\lambda\in\R$  results in a diffracted wavefront, which, after travelling through an assumed homogeneous medium is recorded by the sensor array. 
In a far-field context, the Fraunhofer equation \cite{mast2007fresnel,johnson1992array} permits this wavefront at each sensor position $\bm{p}_i\in\R^3$ to be approximated by:
\begin{align}
Y(\bm{p}_i)&=\int_{\mathbb{S}^2} \,S(\bm{r})\,\mbox{exp}\left(-j\frac{2\pi}{\lambda}\langle\bm{r},\bm{p}_i\rangle\right)\,d\bm{r}\;+\; n_i,
\label{dataModel_basics}
\end{align}
where $i=1,\ldots,L,$ and $\bm{n}=\left[n_1,\ldots,n_L\right]$ is an additive white noise term capturing the inaccuracies in measurement of each sensor, distributed as \cite{krim1996two}
$$ \bm{n}\stackrel{d}{\sim}\mathbb{C}\mathcal{N}_L(\bm{0},\sigma I_L),\quad \sigma >0.$$
Noise across sensors is assumed to be identically and independently distributed and independent of the random amplitude function $\mathcal{S}$. 

We assume that every realisation, or \emph{sample function} \cite{lifshits2012lectures}, $s_\omega:~\mathbb{S}^2\rightarrow \C$ of $\mathcal{S}$ is an element of some Hilbert space $\mathcal{H}=\mathcal{L}^2(\mathbb{S}^2,\C)$ of finite-energy functions, and thus \cref{dataModel_basics} can be written as:
\begin{align*}
Y(\bm{p}_i) &=\langle S,\phi_i\rangle\;+\; n_i,\quad i=1,\ldots,L,
\end{align*}
where $\phi_i(\bm{r}):=\mbox{exp}\left(j2\pi\langle\bm{r},\bm{p}_i\rangle/\lambda\right),$ which in turn
% the space of finite-energy functions, which ensures the integral in \cref{dataModel_basics} to be well-defined. 
can be re-written more compactly using an \emph{analysis operator} \cite{vetterli2014foundations} $\Phi^\ast:\mathcal{H}\rightarrow \C^L$, mapping an element of $\mathcal{H}$ to a finite number $L$ of measurements:
$$\bm{Y}=\left[\begin{array}{c}
Y(\bm{p}_1)\\
\vdots\\
Y(\bm{p}_L)
\end{array}\right] =\left[\begin{array}{c}
\langle S,\phi_1\rangle \\
\vdots\\
\langle S,\phi_L\rangle
\end{array}\right]+\left[\begin{array}{c}
n_1 \\
\vdots\\
n_L
\end{array}\right]=\Phi^\ast S+\bm{n}.$$
We call $\Phi^\ast$ the \emph{sampling operator} \cite{vetterli2014foundations} associated with the sensor array. As the sum of two independent centred complex Gaussian random vectors, the vector of measurements $\bm{Y}$ is also a centred complex Gaussian random vector with covariance matrix $\Sigma\in\C^{L\times L}:$ 
\begin{align}(\Sigma)_{ij}=\iint_{\mathbb{S}^2\times\mathbb{S}^2}\kappa(\bm{r},\bm{\rho}) \,\phi^\ast_i(\bm{r})\phi_j(\bm{\rho})\,d\bm{r}d\bm{\rho}\;+\;\sigma\, \delta_{ij},\label{cov_integral_form}\end{align}
where $i,j=1,\ldots,L,$ $\delta_{ij}$ denotes the Kronecker delta and  $\kappa:\mathbb{S}^2\times \mathbb{S}^2\rightarrow \C$ is the \emph{covariance kernel} \cite{lifshits2012lectures} of $\mathcal{S}$: $$\kappa(\bm{r},\bm{\rho}):=\mathbb{E}\left[S(\bm{r})S^\ast(\bm{\rho})\right],\; \bm{r}, \bm{\rho} \in\mathbb{S}^2.$$
Introducing the associated \emph{covariance operator} $\mathcal{T}_\kappa:\mathcal{H}\rightarrow \mathcal{H}$:
\begin{align*}(\mathcal{T}_\kappa f)(\bm{r}):=\int_{\mathbb{S}^2} \kappa(\bm{r},\bm{\rho})f(\bm{\rho})d\bm{\rho},\quad f\in\mathcal{H},\,\bm{r}\in\mathbb{S}^2,
%\begin{cases}
%\mathcal{H}\rightarrow \mathcal{H}\\
%f\mapsto g(\bm{r})=\int_{\mathbb{S}^2} \kappa(\bm{r},\bm{s})f(\bm{s})d\bm{s},\
%\end{cases}
\end{align*}
we can again reformulate \cref{cov_integral_form} in terms of the sampling operator $\Phi^\ast$ and its adjoint, called the \emph{synthesis operator} \cite{vetterli2014foundations}, $\Phi:\C^L\rightarrow\mathcal{H}$:
\begin{equation}\Sigma=\Phi^\ast \mathcal{T}_{\kappa}\Phi\;+\; \sigma I_L.\label{data_model_visibilities}\end{equation}
By analogy with the finite dimensional case \cite{jinadasa1988applications}, it is customary to write $\kappa=\mbox{vec}(\mathcal{T}_\kappa),$ where the $\mbox{vec}(\cdot)$  operator maps an infinite-dimensional linear operator onto its associated kernel representation. 
Because of the Gaussianity assumption, the covariance kernel $\kappa$ (or equivalently the covariance operator $\mathcal{T}_\kappa$) %can be shown to be a \emph{sufficient statistic} \cite{panaretosstatistics} for $\mathcal{S}$: knowing $\kappa$  fully 
completely determines the distribution of $\mathcal{S}$. Our goal is hence to leverage \cref{data_model_visibilities} in order to form an estimate of $\kappa$ from the covariance matrix $\Sigma$ of the instrument recordings. Often the source field is assumed to be spatially uncorrelated, in which case the random function $\mathcal{S}$ is Gaussian white noise \cite{lifshits2012lectures} and $\kappa$ becomes diagonal. The diagonal part of $\kappa$  $$I(\bm{r}):=\kappa(\bm{r},\bm{r}),\quad \bm{r}\in\mathbb{S}^2, $$ is called the \emph{intensity function}  of the source field, of crucial interest in many  array signal processing applications.
\subsection{Empirical Version}
In practice of course, the covariance matrix $\Sigma$ needs to be estimated from a finite number of i.i.d. observations of $\bm{Y}$, say $N$. Typically, the maximum likelihood estimate of $\Sigma$ is formed by
$\hat{\Sigma}=\frac{1}{N}\sum_{i=1}^N\bm{y}_i\bm{y}_i^H.$
It follows a $L$-variate \emph{complex Wishart distribution} \cite{goodman1963statistical,maiwald2000calculation} with $N$ degrees of freedom and mean $\Sigma$:
\begin{equation}N\hat{\Sigma}\stackrel{d}{\sim}\C\mathcal{W}_L(N,\Sigma).\label{visibilities_distribution}\end{equation}
The density of a complex Wishart distribution can be found in \cite{maiwald2000calculation}. 
In the next section, we use it to form the likelihood function of the data $\hat{\Sigma}$ and derive maximum likelihood estimates of the covariance kernel $\kappa$ and the noise level $\sigma$.
\section{Sieved  Maximum Likelihood}
We now take the population and empirical data models \cref{data_model_visibilities,visibilities_distribution} and derive maximum likelihood estimates for  $\kappa$ and $\sigma $. The simpler case of known noise power, which allows for an insightful geometric interpretation of the maximum likelihood estimate in terms of projection operators, is presented first. That is then followed by the more general case given an unknown noise level.

% and provide maximum likelihood estimates for both the covariance kernel and the noise level. 
\subsection{A Constrained Log-Likelihood Maximisation Problem}
The log-likelihood function for $\kappa$ and $\sigma $ given the sufficient statistic $\hat{\Sigma}$ \cite{panaretosstatistics} can be written in terms of the density function of the complex Wishart distribution \cite{maiwald2000calculation},
{\footnotesize
 \begin{equation}\ell\left(\kappa,\sigma |\hat{\Sigma}\right)=-\mbox{Tr}\left[\left(\Phi^\ast\mathcal{T}_\kappa\Phi+\sigma I_L\right)^{-1}\hat{\Sigma}\right]-\log\left|\Phi^\ast\mathcal{T}_\kappa\Phi+\sigma I_L\right|,\label{loglikelihood_knownnoise}\end{equation}}
 where the terms independent of $\kappa$ and $\sigma$ have been dropped.
%Note that the log-likelihood function is only well-defined for an invertible matrix $\Phi^\ast\mathcal{T}_\kappa\Phi+\sigma I_L$. This condition is fortunately always met in practice, since $\sigma >0$ guarantees invertibility. 
As $\sigma >0$, it is guaranteed that the matrix $\Phi^\ast\mathcal{T}_\kappa\Phi+\sigma I_L$ is invertible and that the log-likelihood function is hence well-defined.
%a condition is fortunately always met in practice, since the noise power plays here the role of a regulariser, making full rank the  potentially rank deficient matrix $\Phi^\ast\mathcal{T}_\kappa\Phi$. 
Maximum likelihood estimates for $\kappa$ and $\sigma $ are then obtained by maximising \cref{loglikelihood_knownnoise} with respect to $\kappa\in\mathcal{L}^2(\mathbb{S}^2\times \mathbb{S}^2)$ and $\sigma >0$.
Since the sampling operator $\Phi^\ast$ has finite rank and consequently a non-trivial kernel, the log-likelihood function admits \emph{infinitely many} local maxima. Indeed, for $f\in\mathcal{N}(\Phi^\ast)$, adding a kernel of the form\footnote{The tensor product $\otimes$ is defined as $\left(\bar{f}\otimes f\right)g:=\langle g,f\rangle f,\quad\forall f,g\in\mathcal{L}^2(\mathbb{S}^2,\C)$.} $\bar{f}\otimes f$ to $\mathcal{T}_\kappa$ in \eqref{loglikelihood_knownnoise} does not change the value of the log-likelihood function. 
%Consider for example a maximising pair $(\hat{\kappa}, \hat{\sigma})$ of \cref{loglikelihood_knownnoise}. Then, the pair $(\tilde{\kappa}, \hat{\sigma})$, where $\tilde{\kappa}=\mbox{vec}\left(\mathcal{T}_{\hat{\kappa}}+\bar{f}\otimes f\right)$,  $f\neq 0$ and $f\in\mathcal{N}(\Phi^\ast)$ is also a maximising pair. Using known properties of the tensor product\footnote{The tensor product $\otimes$ is defined as $\left(\bar{f}\otimes f\right)g:=\langle g,f\rangle f,\quad\forall f,g\in\mathcal{L}^2(\mathbb{S}^2,\C)$.} $\otimes$, we have indeed 
%$\Phi^\ast\left(\mathcal{T}_{\hat{\kappa}}+ \bar{f}\otimes f\right)\Phi%&=\Phi^\ast\mathcal{T}_{\hat{\kappa}}\Phi+ \Phi^\ast\left(\bar{f}\otimes f\right)\Phi\\
%%=\Phi^\ast\mathcal{T}_{\hat{\kappa}}\Phi+\left(\bar{\Phi}^\ast\otimes\Phi^\ast\right)\left(\bar{f}\otimes f\right)
%=\Phi^\ast\mathcal{T}_{\hat{\kappa}}\Phi+\overline{\Phi^\ast f}\otimes\Phi^\ast f=\Phi^\ast\mathcal{T}_{\hat{\kappa}}\Phi,$
%and hence plugging $(\tilde{\kappa}, \hat{\sigma})$ in \cref{loglikelihood_knownnoise} yields the same log-likelihood value as for the assumed maximising pair $(\hat{\kappa}, \hat{\sigma})$. 
We thus choose to impose a unique maximum by restricting the search space for $\kappa$ to a lower dimensional subspace, and look for solutions in the range of some synthesis operator $\bar{\Psi}\otimes \Psi$, which will be specified in \cref{choice_psi}:
%In order for \cref{loglikelihood_knownnoise} to admit a unique maximum, we must hence restrict  the search space for $\kappa$ to a lower dimensional subspace, and look for solutions in the range of some synthesis operator $\bar{\Psi}\otimes \Psi$:
\begin{equation*}\kappa=\left(\bar{\Psi}\otimes\Psi\right)\mbox{vec}(R)=\sum_{i,j=1}^M R_{ij} \;\bar{\psi}_j\otimes\psi_i,\;\Leftrightarrow\; \mathcal{T}_\kappa=\Psi R\Psi^\ast,\label{assumption_func}\end{equation*}
 where $R\in\C^{M\times M}$ is a Hermitian symmetric matrix and $\Psi^\ast:\mathcal{H}\rightarrow \C^M$, $\Psi:\C^M\rightarrow \mathcal{H}$ are the analysis and synthesis operators associated with the family of functions $\{\psi_1,\ldots,\psi_M\}\subset\mathcal{H}$. This regularisation of the likelihood problem by restricting the parameter space to a lower dimensional subspace is generally known as the \emph{method of sieves} \cite{grenander1981abstract,geman1982nonparametric}. The maximum likelihood estimates of $R$ and $\sigma $ are then given by minimising the negative log-likelihood:
{\footnotesize\begin{equation}\hat{R}, \hat{\sigma}=\mbox{arg}\min_{\substack{R\in\C^{M^2}\\\sigma >0}}\mbox{Tr}\left[\left(G R G^H+\sigma I_L\right)^{-1}\hat{\Sigma}\right]+\log\left|G RG^H+\sigma I_L\right|,\label{ml_opt_joint}\end{equation}}
where $G=\Phi^\ast\Psi\in\C^{L\times M}$ is the so-called \emph{Gram matrix} \cite{vetterli2014foundations}, given by $(G)_{ij}=\langle\psi_j,\phi_i\rangle$. For \cref{ml_opt_joint} to admit a unique solution, it is necessary to have at least as many measurements as unknowns. When the noise power is unknown a priori, this requires that $M<L$. When the noise power is known, there is one less unknown, leading to $M\leq L$. This is however not a sufficient condition for identifiability, and we must further assume $G$ to be of full column-rank. If the latter condition is verified, we say that the two families of functions $\{\phi_1,\ldots,\phi_L\}$ and $\{\psi_1,\ldots,\psi_M\}$ are \emph{coherent} with one another.

 \subsection{Estimation with Known Noise Power}
\label{section_known_noise}
Suppose the noise power $\sigma $ is known. Then $R$ becomes the only variable in \cref{ml_opt_joint}, and a solution can easily be obtained by cancelling the derivative. This yields 
$$\hat{R}=G^\dagger\tilde{\Sigma}\left(G^\dagger\right)^H=G^\dagger\left[\hat{\Sigma}-\sigma I_L\right]\left(G^\dagger\right)^H,$$
where $G^\dagger\in\C^{L\times M}$ is the \emph{left pseudo-inverse}\footnote{The left pseudo-inverse exists since $G$ is assumed full-column rank.} \cite{engl1996regularization} of $G$. Hence, when restricting the search space to $\mathcal{R}(\bar{\Psi}\otimes\Psi)$, the maximum likelihood of $\kappa$ is given by
\begin{align}\hat{\kappa}&=\sum_{i,j=1}^M \hat{R}_{ij} \;\bar{\psi}_j\otimes\psi_i\nonumber\\
%&=\left(\bar{\Psi}\otimes\Psi\right)\mbox{vec}(\hat{R})\nonumber\\
&=\left(\bar{\Psi}\otimes\Psi\right)\mbox{vec}\left(G^\dagger\left[\hat{\Sigma}-\sigma I_L\right]\left(G^\dagger\right)^H\right)\nonumber\\
&=\left(\bar{\Psi}\otimes\Psi\right)\left[\bar{G}^\dagger\otimes G^\dagger\right]\left(\hat{\bm{\varsigma}}-\sigma \bm{\epsilon}\right),\label{estimate_kappa}
\end{align}
 with $\hat{\bm{\varsigma}}=\mbox{vec}(\hat{\Sigma})\in\C^{L^2}$ and $\bm{\epsilon}=\mbox{vec}(I_L)\in\C^{L^2}.$ The intensity function is then obtained by taking the diagonal part of $\hat{\kappa}:$
 $$\hat{I}(\bm{r})=\sum_{i,j=1}^M\hat{R}_{ij}\psi_i(\bm{r})\bar{\psi}_j(\bm{r}),\quad\forall\bm{r}\in\mathbb{S}^2.$$
 Using properties of the tensor product and the vec operator, we can re-write \cref{data_model_visibilities} as
 $ \bm{\varsigma}=\left(\bar{\Phi}\otimes\Phi\right)^\ast\kappa \;+\;\sigma \bm{\epsilon}.$
 Hence, since $\mathbb{E}[\hat{\bm{\varsigma}}]=~\bm{\varsigma}$, \cref{estimate_kappa} becomes on expectation 
 $$\mathbb{E}[\hat{\kappa}]=\left(\bar{\Psi}\otimes\Psi\right)\left[\bar{G}^\dagger\otimes G^\dagger\right]\left(\bar{\Phi}\otimes\Phi\right)^\ast\kappa.$$
For $M=L$, $G$ is invertible and $G^\dagger=G^{-1}$, making  $(\bar{\Psi}\otimes\Psi)[\bar{G}^{-1}\otimes G^{-1}](\bar{\Phi}\otimes\Phi)^\ast$ an \emph{oblique projection} operator \cite{vetterli2014foundations}. The operator $(\bar{\Psi}\otimes\Psi)[\bar{G}^{-1}\otimes G^{-1}]$ is indeed a \emph{right-inverse} of  $ (\bar{\Phi}\otimes\Phi)^\ast$:
\begin{equation} \left(\bar{\Phi}\otimes\Phi\right)^\ast\left(\bar{\Psi}\otimes\Psi\right)\left[\bar{G}^{-1}\otimes G^{-1}\right]=\overline{\Phi^\ast\Psi G^{-1}}\otimes\Phi^\ast\Psi G^{-1}=I_{L^2}.\label{consistency}\end{equation}
In the specific case where $M=L$, the maximum likelihood estimate $\hat{\kappa}$ is hence an unbiased, consistent and asymptotically efficient estimator of the \emph{oblique projection} of $\kappa$ onto $\mathcal{R}(\bar{\Psi}\otimes\Psi)$. When additionally setting  $\Psi=\Phi$ the projection becomes orthogonal. 
%$$\mathbb{E}[\hat{\kappa}]=\mbox{arg}\left\{\min_{\chi\in\mathcal{R}(\bar{\Phi}\otimes\Phi)} \|\chi-\kappa\|_2^2\right\}.$$
\subsection{Joint Estimation}
\label{estimation_unkown_noise}
%The log-likelihood function for $\kappa$ and $\sigma $ given the data $\hat{\Sigma}$ is given by 
%{\footnotesize
% \begin{equation}\ell\left(\kappa,\sigma |\hat{\Sigma}\right)=-\mbox{Tr}\left[\left(\Phi^\ast\mathcal{T}_\kappa\Phi+\sigma I_L\right)^{-1}\hat{\Sigma}\right]-\log\left|\Phi^\ast\mathcal{T}_\kappa\Phi+\sigma I_L\right|,\label{loglikelihood_unknownnoise}\end{equation}}
%Note again that the log-likelihood function is only well-defined for an invertible matrix $\Phi^\ast\mathcal{T}_\kappa\Phi+\sigma I_L$. This condition is however easily met in practice, since the noise power plays here the role of a regulariser, adding a small amount of rank\footnote{as a matter of fact $\sigma \geq1$ is sufficient to guarantee invertibility.} to a potentially rank deficient matrix $\Phi^\ast\mathcal{T}_\kappa\Phi$. To obtain a unique maximiser of \cref{loglikelihood_unknownnoise}, we constrain the search space as in \cref{assumption_func}. Maximum likelihood estimates for  $R$ and $\sigma $ are then given by minimising the negative log-likelihood function:
%{\footnotesize\begin{equation}\hat{R}, \hat{\sigma}=\mbox{arg}\left\{\min_{R,\sigma }\mbox{Tr}\left[\left(G R G^H+\sigma I\right)^{-1}\hat{\Sigma}\right]+\log\left|G RG^H+\sigma I\right| \right\}.\label{ml_opt_joint}\end{equation}}
%Since we have one more parameter to estimate, identifiability is this time guaranteed for $M<L$ and $G$ full column-rank. 
Suppose now the noise power is unknown. We must hence minimise \cref{ml_opt_joint} with respect to both $\sigma $ and $R$. Using the result from theorem 1.1 of \cite{stoica1995concentrated}, we can write explicit solutions for the unique minimisers of \cref{ml_opt_joint}:
\begin{align} \hat{\sigma}=\frac{\mbox{Tr}\left(\hat{\Sigma}-GG^\dagger \hat{\Sigma}\right)}{L-M},\qquad \hat{R}=G^\dagger \left[\hat{\Sigma}- \hat{\sigma} I\right]{\left(G^\dagger\right)}^H.\end{align}
Again, the constrained maximum likelihood estimate of $\kappa$ is given by 
\begin{equation}\hat{\kappa}=\left(\bar{\Psi}\otimes\Psi\right)\left[\bar{G}^\dagger\otimes G^\dagger\right]\left(\hat{\bm{\varsigma}}- \hat{\sigma}\bm{\epsilon}\right),\label{estimate_ml_unknwon}\end{equation}
with intensity function $\hat{I}(\bm{r})=\sum_{i,j=1}^M\hat{R}_{ij}\psi_i(\bm{r})\bar{\psi}_j(\bm{r}).$ This time, since $M<L$ the consistency condition \cref{consistency} cannot be met, and $\mathbb{E}[\hat{\kappa}]$ can no longer be interpreted as an oblique projection of $\kappa$. For values of $M$ comparable to $L$ though, the consistency condition should still hold approximately\footnote{More precisely, the consistency condition will hold on a subspace of $\C^L$ of dimension $M$.}
, and this geometrical interpretation provides intuition.
\subsection{On the choice of $\Psi$}
\label{choice_psi}
\begin{figure}[p!]
\begin{center}
\subfloat[][BIC profile for the setup described in \cref{comparison_ SiML_vssrp}.]{
\includegraphics[width=0.46\linewidth,valign=b]{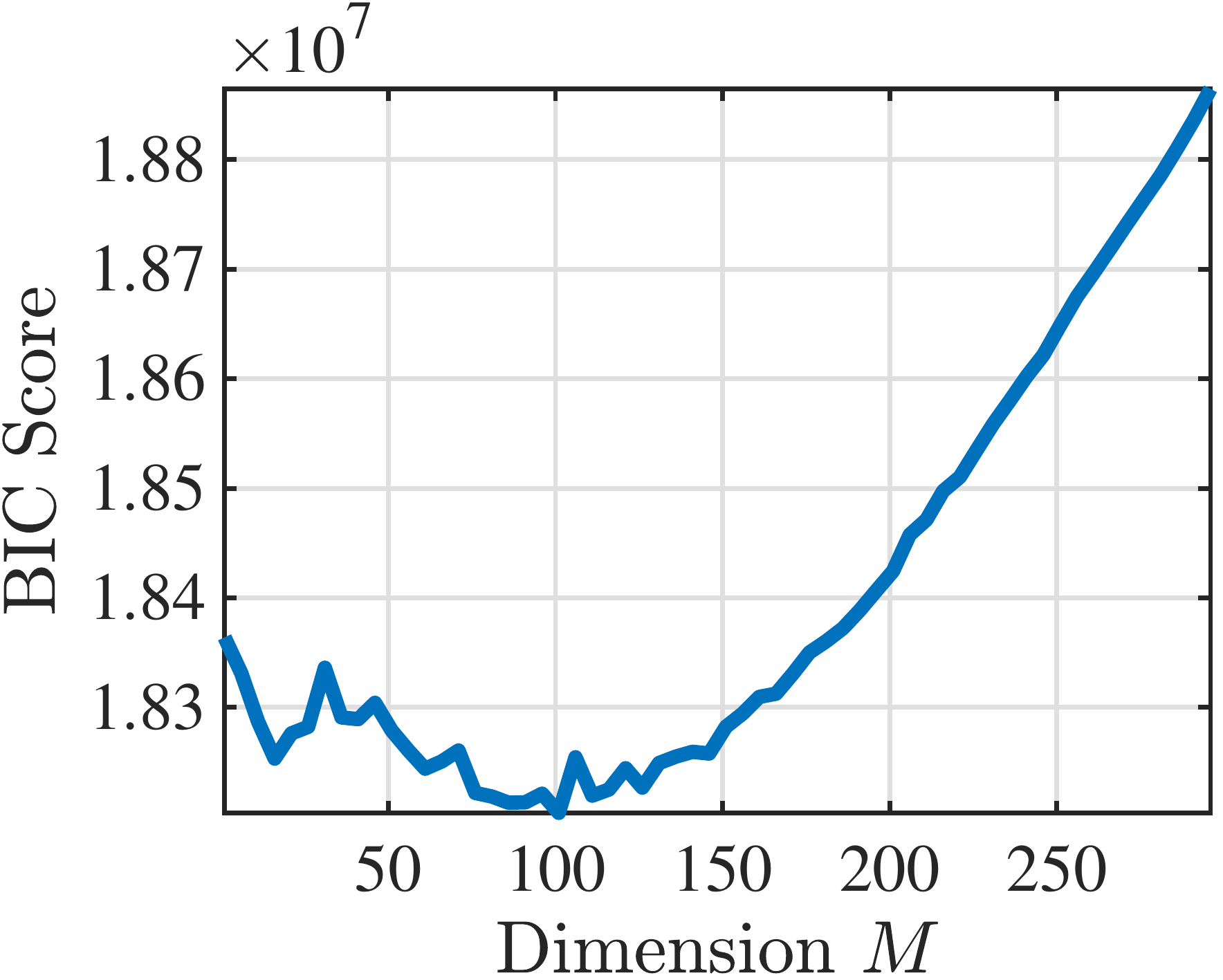}
}ß
\subfloat[][BIC-selected $M$ vs. SNR for the setup described in \cref{comparison_ SiML_vssrp}.]{
\includegraphics[width=0.46\linewidth,valign=b]{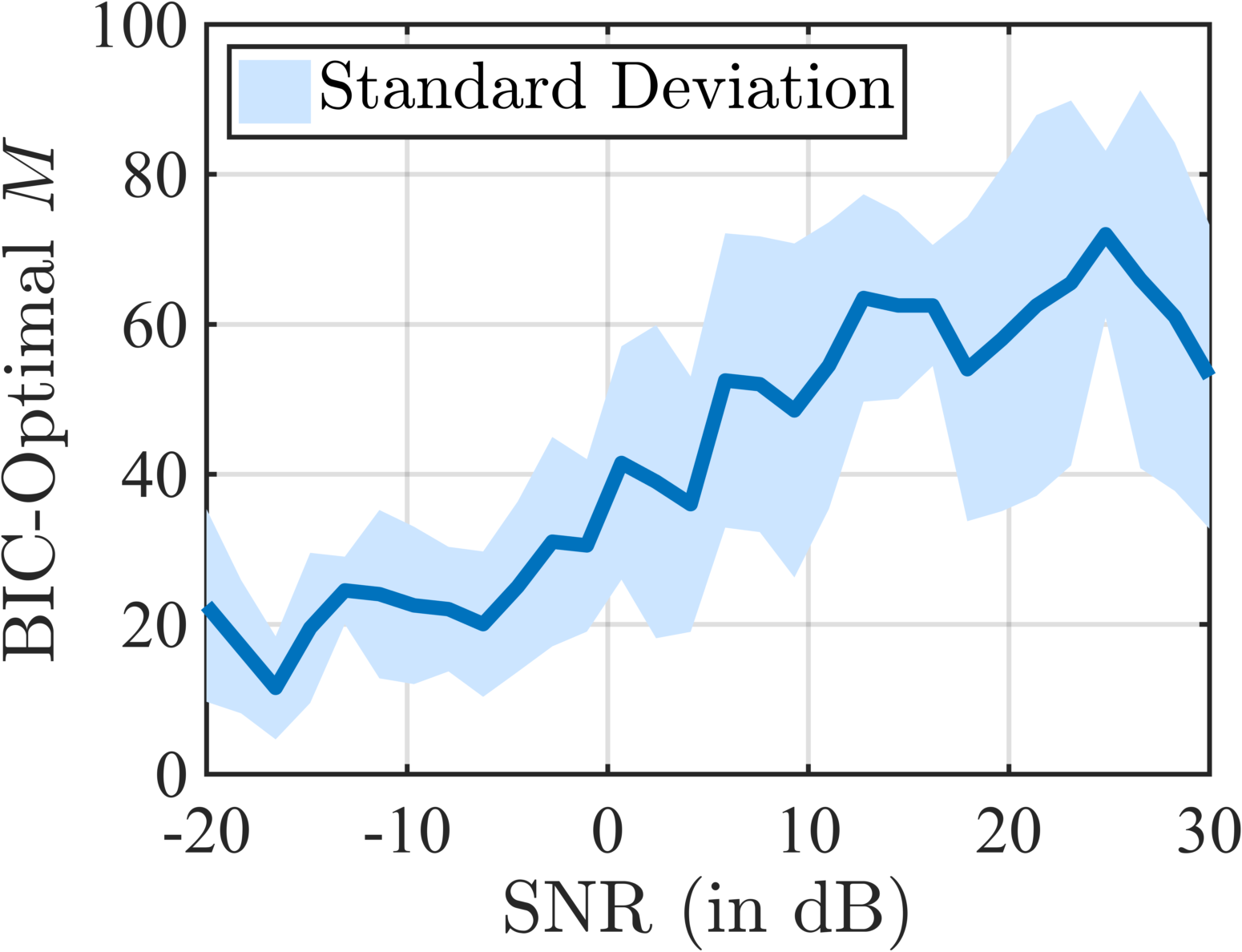}
}\vspace{-1em}
\end{center}
\caption{Optimal choice of the dimension $M$ based on the Bayesian Information Criterion (BIC).}
\label{BIC_choice}
\end{figure}

\begin{figure}[p!]
\begin{center}
\subfloat[][Intensity field.]{
\includegraphics[width=0.3\linewidth,valign=c]{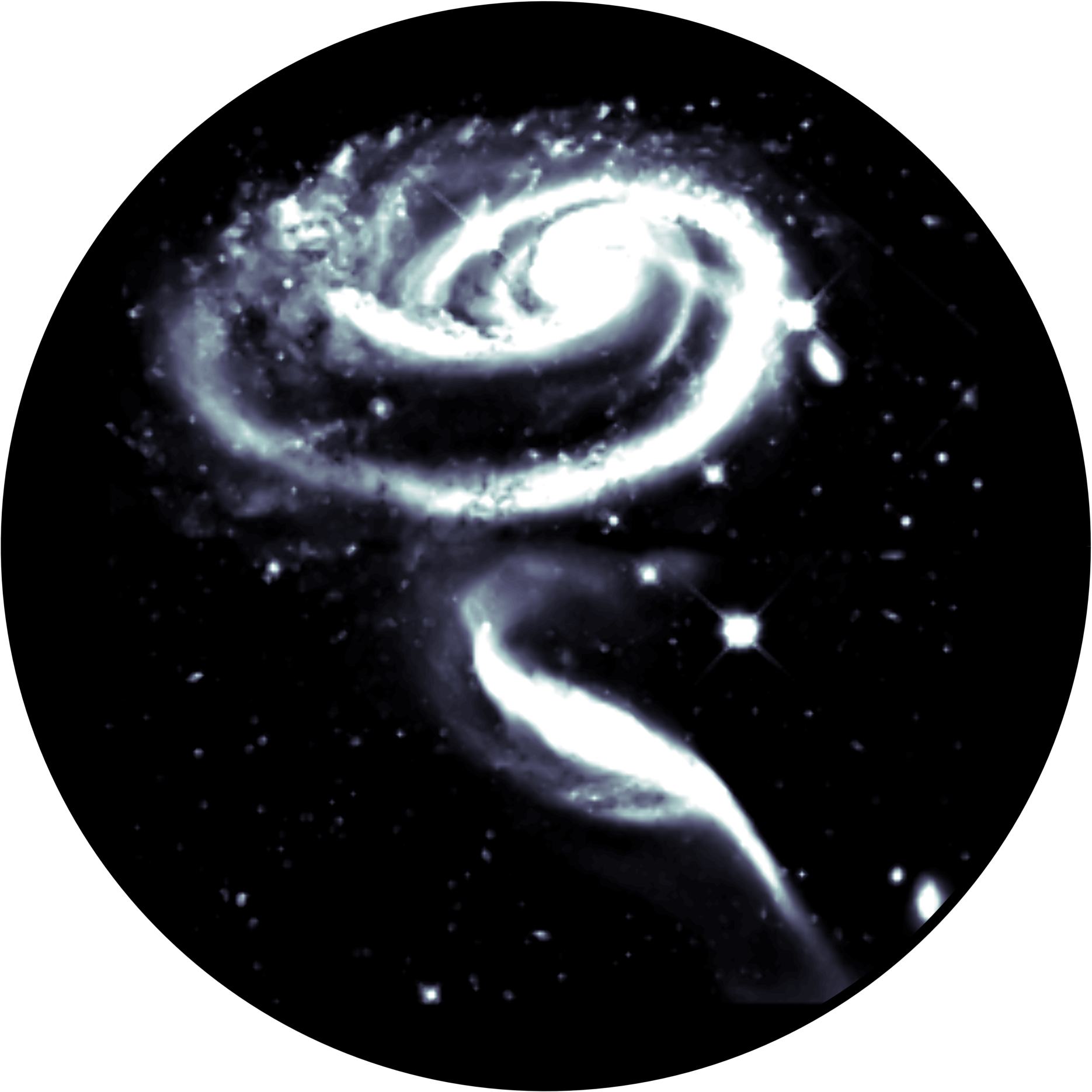}
\label{intensity_field}
}\;
\subfloat[][Relative MSE and RMS contrast scores of the various images below.]{
\includegraphics[width=0.62\linewidth,valign=c]{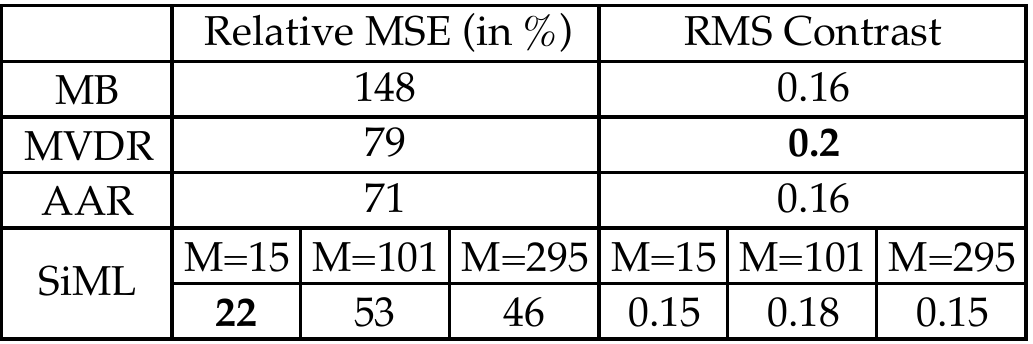}
}

\subfloat[][MB estimate.]{
\includegraphics[width=0.3\linewidth]{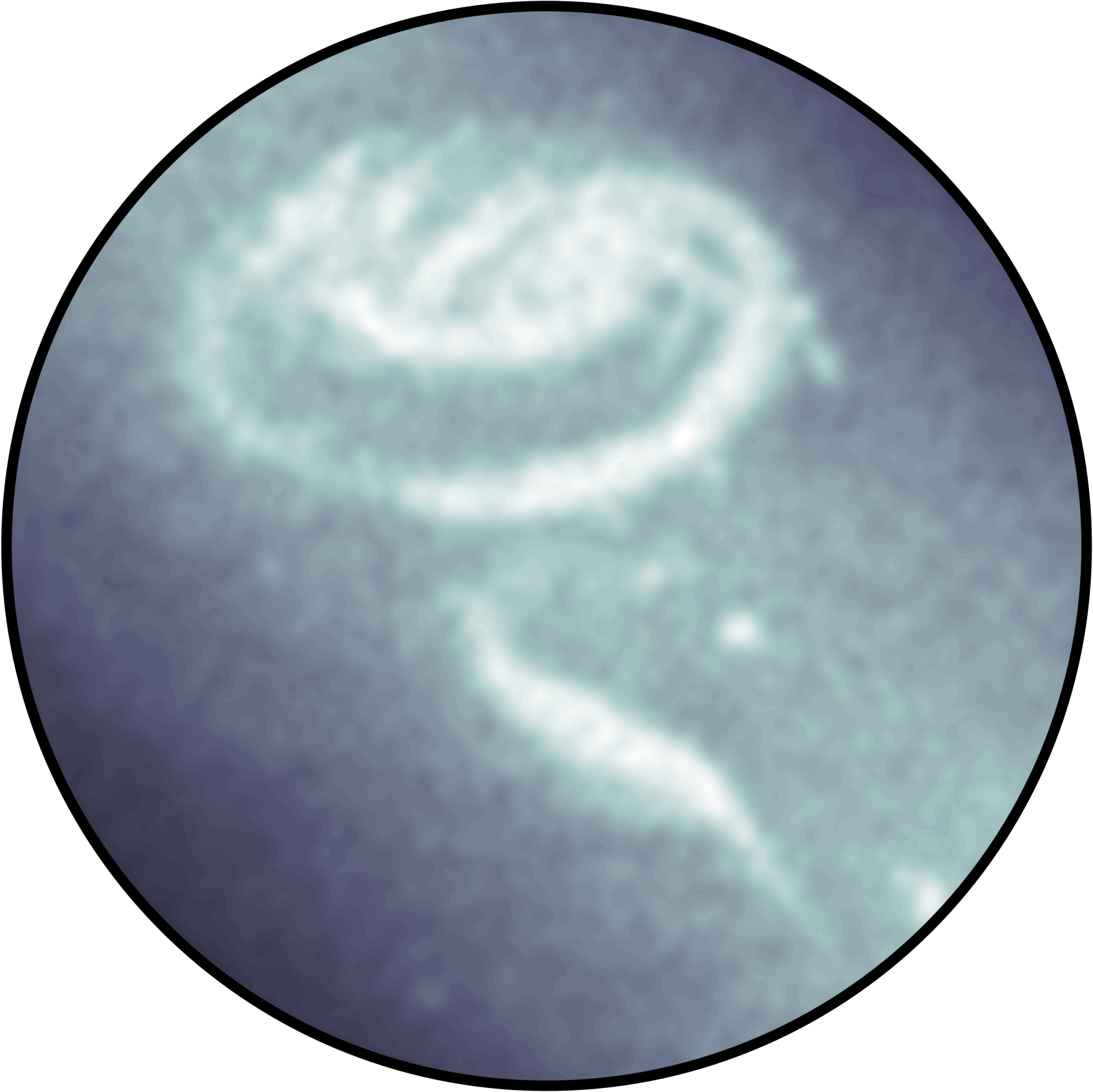}
}
\subfloat[][MVDR estimate.]{
\includegraphics[width=0.3\linewidth]{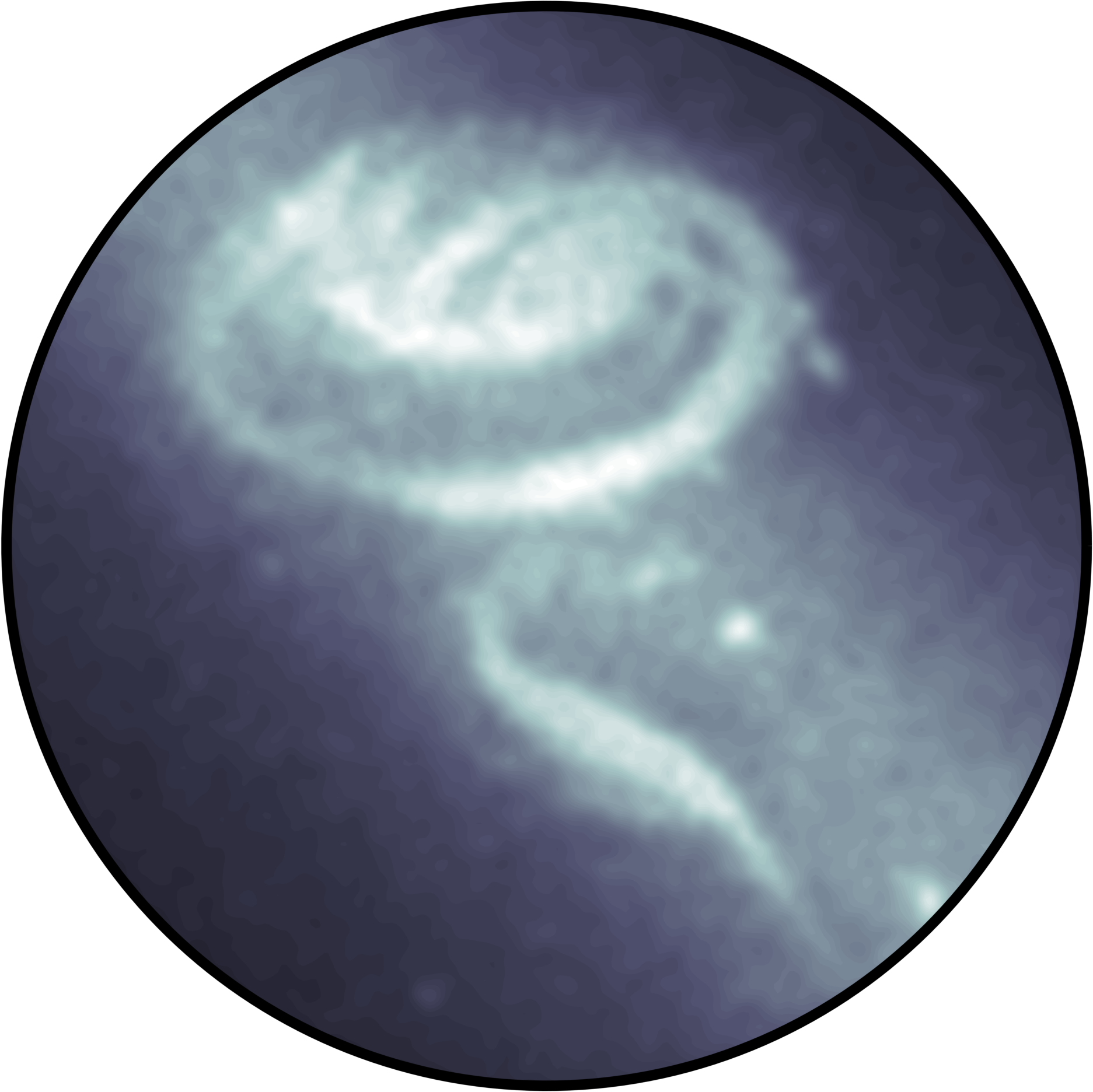}
}
\subfloat[][AAR estimate.]{
\includegraphics[width=0.3\linewidth]{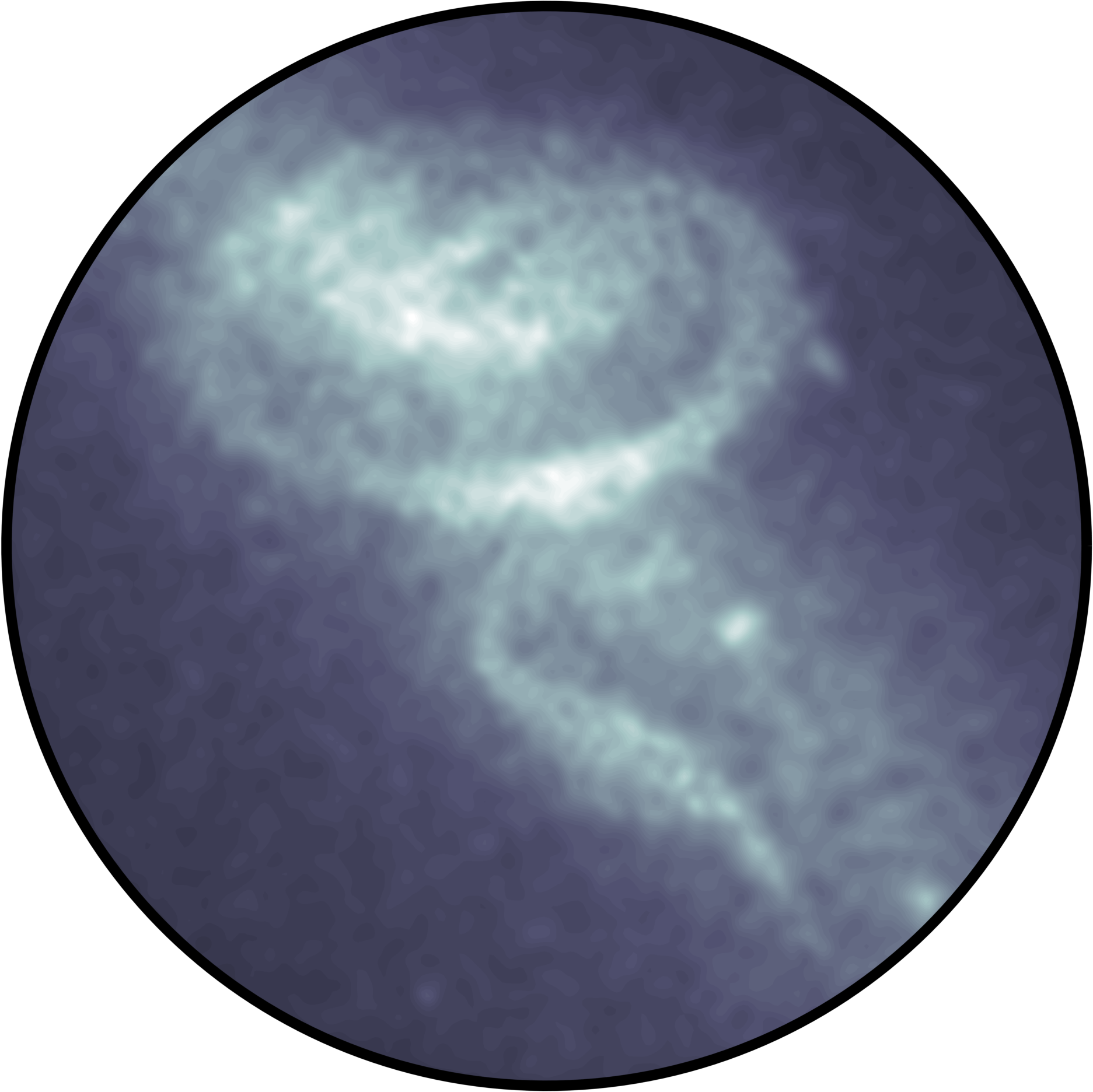}
}

%\subfloat[][MB absolute error.]{
%\includegraphics[width=0.3\linewidth]{mb_error_sml.png}
%}\;
%\subfloat[][MVDR absolute error.]{
%\includegraphics[width=0.3\linewidth]{mvdr_error_sml.png}
%}\;
%\subfloat[][AAR absolute error.]{
%\includegraphics[width=0.3\linewidth]{aar_error_sml.png}
%}

\subfloat[][ SiML estimate\\ ($M=15$).]{
\includegraphics[width=0.3\linewidth]{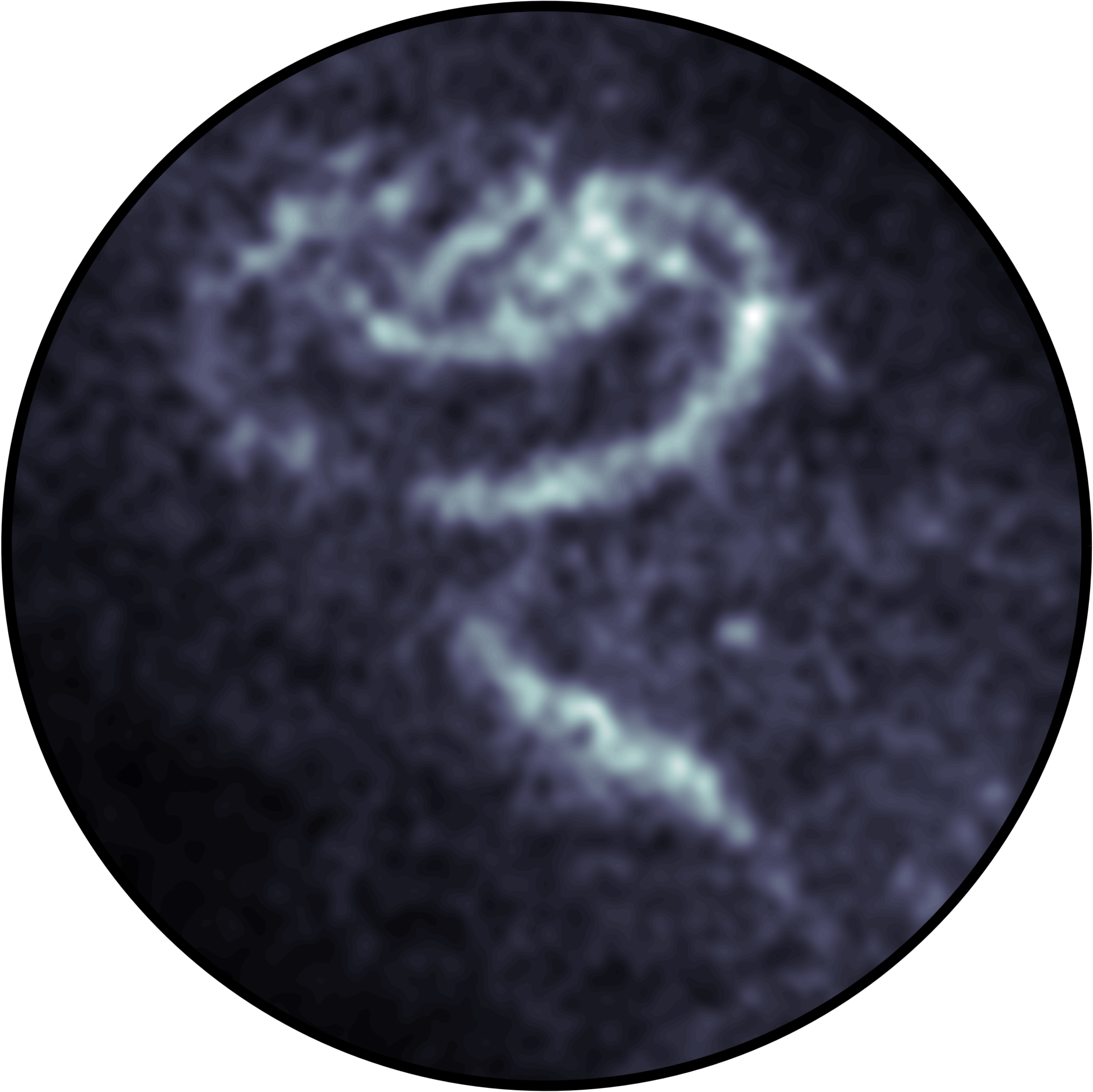}
\label{sml15}
}
\subfloat[][ SiML estimate\\ (BIC-selected\\ $M=101$).]{
\includegraphics[width=0.3\linewidth]{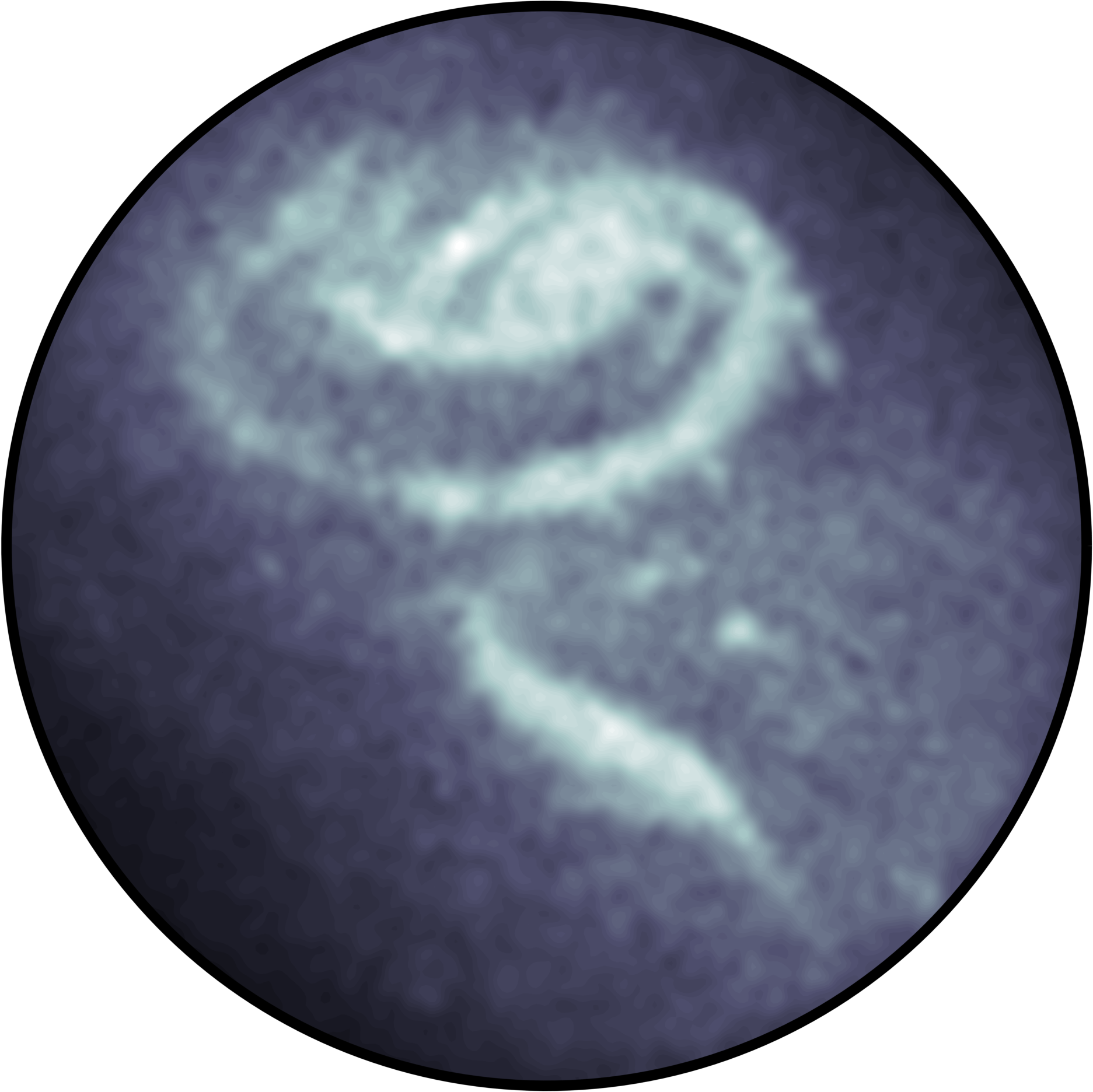}
\label{sml101}
}
\subfloat[][ SiML estimate\\ ($M=296$).]{
\includegraphics[width=0.3\linewidth]{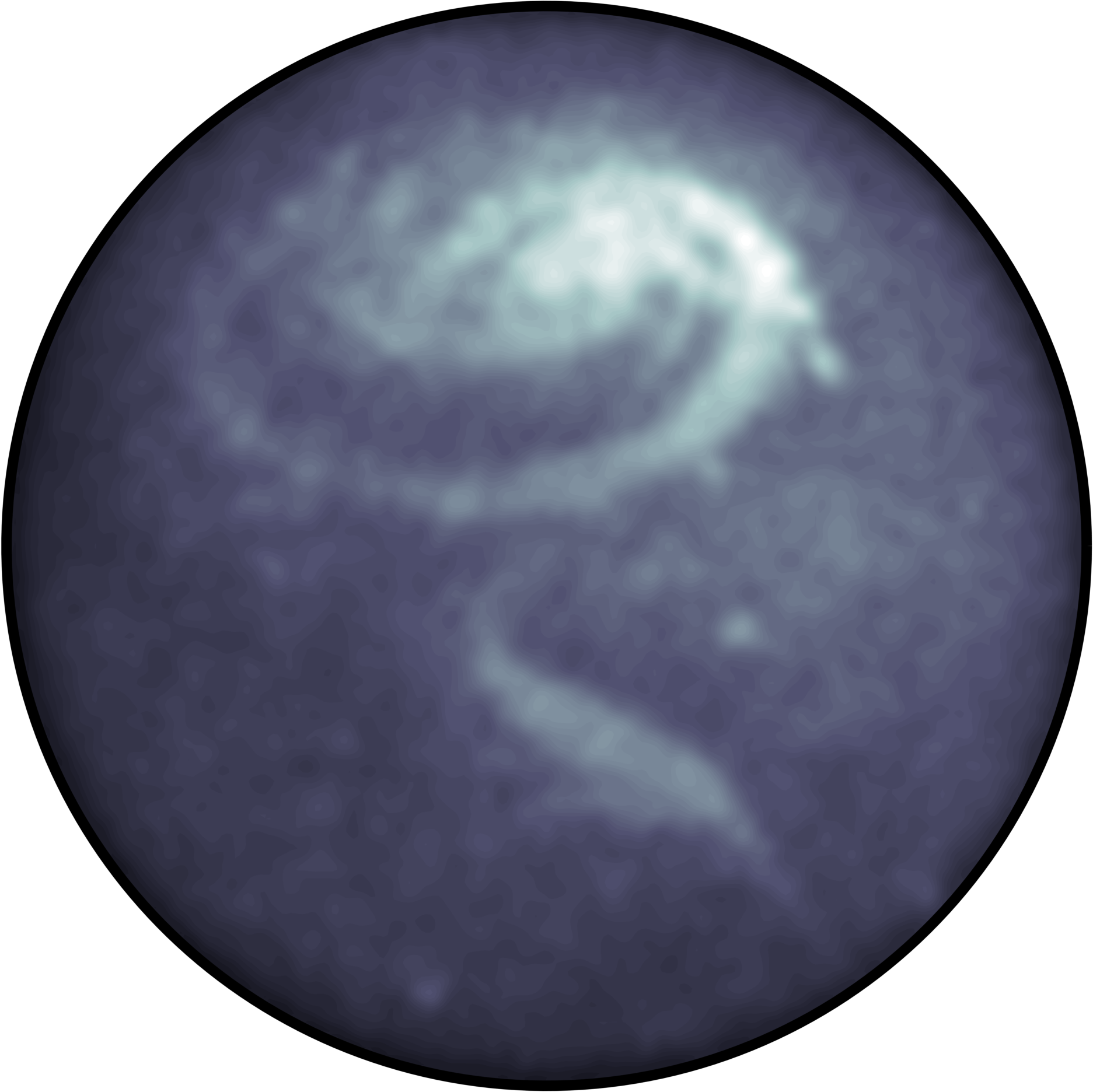}
\label{sml296}
}

\subfloat[][Performance of the different algorithms for various SNR. The standard deviation was obtained over 10 repeated experiments.]{
\includegraphics[width=\linewidth,valign=c]{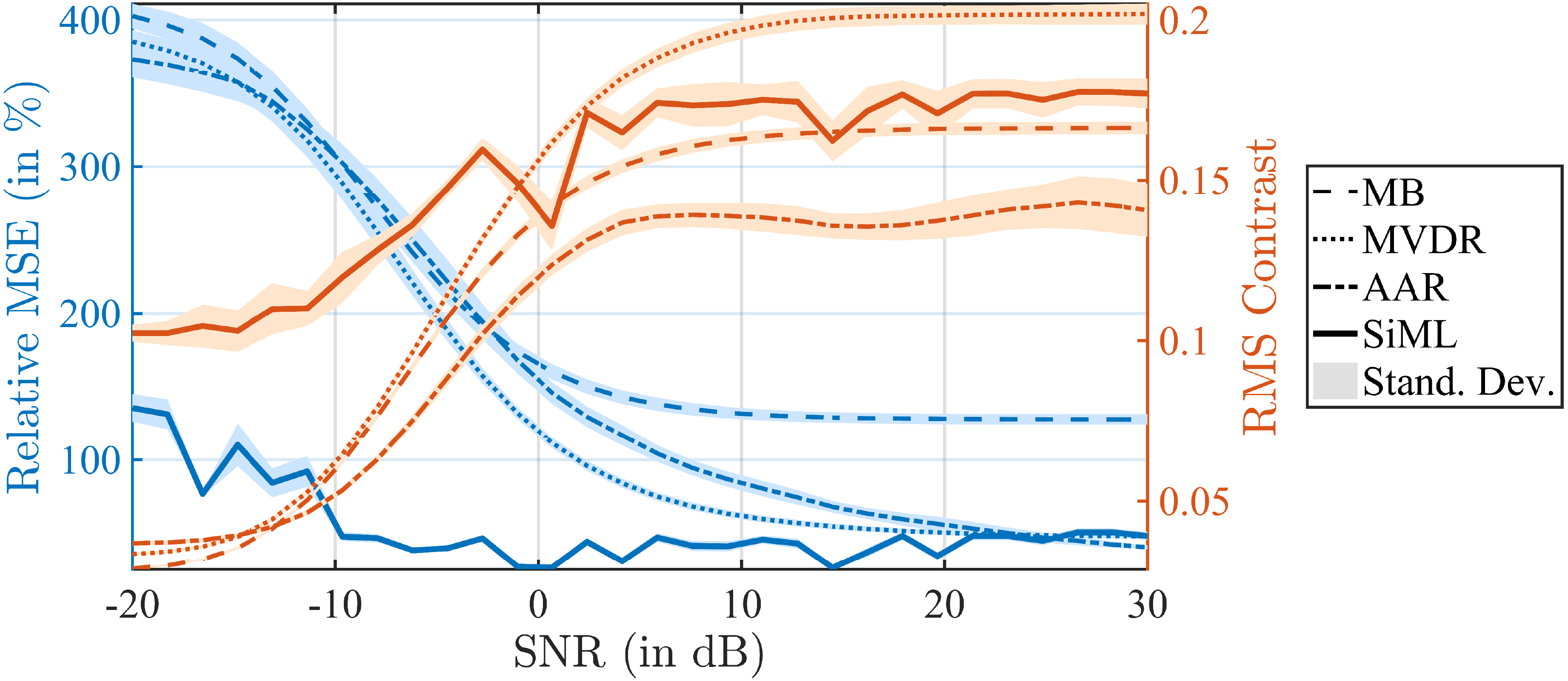}
}\vspace{-1em}
%\subfloat[][SML absolute error ($M=20$).]{
%\includegraphics[width=0.3\linewidth]{ml20_error_sml.png}
%}\;
%\subfloat[][SML absolute error ($M=100$).]{
%\includegraphics[width=0.3\linewidth]{ml100_error_sml.png}
%}\;
%\subfloat[][SML absolute error ($M=299$).]{
%\includegraphics[width=0.3\linewidth]{ml299_error_sml.png}
%}
\end{center}
\caption{Comparison between sieved maximum likelihood method and various steered-response power methods (MB, MVDR, AAR). The parameters of the experiment were set to: $L=300$, $N=2000$, $SNR=5\,\mbox{dB}$.}
\label{comparison_ SiML_vssrp}
\end{figure}

We have thus far only required the synthesis operator $\Psi$ to be identifiable, with the coherency condition requiring  $G=\Phi^\ast\Psi$ to be full column-rank. This still leaves plenty of potential candidates. For practical purposes, we recommend taking  $\Psi=\Phi W$ where $W\in\C^{L\times M}$ is a tall matrix, with columns containing the first $M$ eigenvectors of $\hat{\Sigma}$ (assuming  eigenvalues sorted in descending order). Such a choice presents numerous advantages. First, since $\mathcal{R}(\Phi)^\perp=\mathcal{N}(\Phi^\ast)$, the instrument can only sense functions within the range of $\Phi$, and it is hence natural to choose $\mathcal{R}(\Psi)=\mathcal{R}(\Phi)$. This canonical choice moreover yields an analytically computable Gram matrix $G$. Indeed, we have $G=\Phi^\ast\Phi W=HW$, where $H\in\C^{L\times L}$ is given by (see of \cite[Chapter~4 section 1.1]{simeoni2015towards}):
$$(H)_{ij}=~4\pi\;\mbox{sinc}(2\pi\|\bm{p}_i-\bm{p}_j\|_2/\lambda),\quad i,j=1,\ldots,L.$$
Finally, by choosing the columns of $W$ as the first $M$ eigenvectors of $\hat{\Sigma}$, $M$ acts as a regularisation parameter. Indeed, the eigenvectors associated to the smallest eigenvalues of $\hat{\Sigma}$ are usually the most polluted by noise. Hence, truncating to the $M$ largest eigenvalues reduces the contribution of the noise in the final estimate (see  \cref{sml15,sml101,sml296}). Moreover, small values of $M$ will increase the chances of $(G^HG)\in\C^{M\times M}$ in the left pseudo-inverse  $G^\dagger=(G^HG)^{-1}G^H$ being well-conditioned, thus improving the overall numerical stability of the algorithm.   Suitable values of $M$ can be obtained by minimising the \emph{Bayesian Information Criterion (BIC)} \cite{bhat2010derivation}, often used in model selection:
$ BIC(M)=-2\hat{\ell}_M+2M^2\log(L),$
where $\hat{\ell}_M$ is the maximised log-likelihood function for a specific choice of $M$. Example of a BIC profile and evolution of the BIC-selected $M$ with the signal-to-noise ratio are depicted in \cref{BIC_choice}. \vspace{-0.5em}
  % The dimension $M$ of the matrix $R$ can further be chosen optimally using a statistical test. Indeed, for $N\rightarrow \infty$, we have \cite{ottersten1998covariance} $\sqrt{N}(\hat{\bm{\varsigma}}-~\bm{\varsigma})\stackrel{d}{\sim}\C\mathcal{N}_{L^2}(\bm{0},\bar{\Sigma}\otimes \Sigma)$ and hence
%\begin{align}
%\chi&=N\left\lVert\bar{\Sigma}^{-1/2}\otimes \Sigma^{-1/2}(\hat{\bm{\varsigma}}-\bm{\varsigma})\right\rVert^2_2\nonumber\\
%&=N\left\lVert \Sigma^{-1/2}\hat{\Sigma}\Sigma^{-1/2}-I_L\right\rVert^2_F\;\stackrel{d}{\sim}\;\chi^2\left(L^2\right),
%\label{stat_test}
%\end{align}
%where $\|\cdot\|_F$ denotes the Frobenius norm and $\chi^2(L^2)$ denotes the chi-squared distribution with $L^2$ degrees of freedom. For a given value of $M$, we can then test the null hypothesis $H_0: \Sigma=G\hat{R}G^H+ \hat{\sigma}I_L$ using \cref{stat_test}. Following Occam's razor principle, we then select the smallest value of $M$ for which the hypothesis $H_0$ cannot be rejected for the chosen significance level. 
\subsection{Simulation Results}
\cref{comparison_ SiML_vssrp} compares the performance of the proposed \emph{Sieved Maximum Likelihood (SiML)} method in a radio astronomy setup to three popular spectral-based methods, namely \emph{Matched Beamforming (MB)}, \emph{Maximum Variance Distortionless Response (MVDR)} and  the \emph{Adapted Angular Response (AAR)} \cite{van2013signal}. For this experiment, we generated randomly a layout $L=300$ antennas and simulated $N=2000$ random measurements from the ground truth intensity field \cref{intensity_field}. Furthermore, we considered two metrics to assess the quality of the recovered images: the traditional relative Mean Squared Error (MSE) and the Root Mean Squared (RMS) metric, which measures the contrast of an image by computing its standard deviation over all pixels. The simulations reveal that the  SiML outperforms all the traditional algorithms for the considered SNR range in both metrics, except for large SNRs where MVDR exhibits a slightly better contrast. As for the traditional SML method,  SiML performs particularly well for challenging scenarios with very low SNR.  \vspace{-1em}
%Computationally speaking finally, the tensor product structure in \cref{estimate_ml_unknwon} make the estimate $\hat{\kappa}$ very efficient to compute. This is in total contrast with traditional SML, which requires minimising a highly non-linear $Q$-dimensional function, where $Q<L$ is the number of point sources \cite{krim1996two}.
\section{Conclusion}
SiML generalises the traditional SML method to a wider class of signals, encompassing arbitrarily shaped, possibly correlated,  sources of which there may be more than the number of sensors. The method is numerically stable and admits a nice geometrical interpretation in the case of known noise power. Simulations revealed its superiority with respect to state-of-the-art subspace-based methods, both in terms of accuracy and contrast. Finally, the tensor product structure in \cref{estimate_ml_unknwon} makes the estimate $\hat{\kappa}$ very efficient to compute. This is in  contrast to traditional SML, which requires minimising a highly non-linear multi-dimensional function \cite{krim1996two}.


  \cleardoublepage
\bibliographystyle{IEEEbib}
\bibliography{siml.bib}

\end{document}